\documentclass[sigconf]{acmart}

\usepackage{graphicx}
\usepackage{todonotes}

\usepackage{acronym}

\acrodef{IR}{Information Retrieval}
\acrodef{MLP}{multilayer perceptron}
\acrodef{SERP}{Search Engine Result Page}
\acrodef{ERP}{Event Related Potential}
\acrodef{EEG}{electroencephalogram}
\acrodef{GBDT}{Gradient Boosting Decision Tree}
\acrodef{SST}{SST-EmotionNet}
\acrodef{AUC}{Area Under Curve}
\acrodef{IN}{Information Need}
\acrodef{fMRI}{functional Magnetic Resonance Imaging}
\acrodef{BMI}{Brain–Machine Interface}
\acrodef{BP}{Band Power}
\acrodef{DE}{differential entropy}
\acrodef{RASM}{rational asymmetry}
\acrodef{DASM}{differential asymmetry}
\acrodef{SVM}{Support Vector Machines}
\acrodef{DBN}{Deep Belief Networks}
\acrodef{KNN}{k-Nearest Neighbors}
\acrodef{HBG}{Height-Biased Gain}

\acrodef{BMSI}{Brain Machine Search Interface}
\acrodef{SSVEP}{Steady-State Visually Evoked Potential}
\acrodef{CCA}{Canonical Correlation Analysis}
\AtBeginDocument{%
  \providecommand\BibTeX{{%
    \normalfont B\kern-0.5em{\scshape i\kern-0.25em b}\kern-0.8em\TeX}}}

\AtBeginDocument{%
  \providecommand\BibTeX{{%
    \normalfont B\kern-0.5em{\scshape i\kern-0.25em b}\kern-0.8em\TeX}}}

\acmConference[WSDM '22]{WSDM '22: ACM International Conference on Web Search and Data Mining}{February 21--25 2022}{Phoenix, AZ, USA}
\acmBooktitle{WSDM '22: Web Search via an Efficient and Effective Brain-Machine Interface, February 21--25, 2022, Phoenix, AZ, USA}
\acmPrice{15.00}
\acmISBN{978-1-4503-XXXX-X/18/06}

\begin{document}

\title{Web Search via an Efficient and Effective Brain-Machine Interface}

\author{Xuesong Chen$^{1}$, Ziyi Ye$^{1}$, Xiaohui Xie$^{1}$, Yiqun Liu$^{1\star}$,  Weihang Su$^{2}$, Shuqi Zhu$^{1}$, Min Zhang$^{1}$, Shaoping Ma$^{1}$}
\affiliation{%
  \institution{${1}$ Department of Computer Science and Technology, Institute for Artificial Intelligence, Beijing National Research Center for Information Science and Technology, Tsinghua University}
 \city{Beijing}
 \country{China}
  }
\affiliation{%
  \institution{${2}$ School of Computer Science, Beijing University of Posts and Telecommunications}
 \city{Beijing}
 \country{China}
  }
\email{yiqunliu@tsinghua.edu.cn}

\renewcommand{\shortauthors}{Chen, et al.}
\begin{abstract}
  While search technologies have evolved to be robust and ubiquitous, the fundamental interaction paradigm has remained relatively stable for decades. 
  With the maturity of \ac{BMI}, we build an efficient and effective communication system between human beings and search engines based on electroencephalogram~(EEG) signals, called \ac{BMSI} system.
  The \ac{BMSI} system provides functions including query reformulation and search result interaction. 
  In our system, users can perform search tasks without having to use the mouse and keyboard.
  Therefore, it is useful for application scenarios in which hand-based interactions are infeasible, e.g, for users with severe neuromuscular disorders.
  Besides, based on brain signals decoding, our system can provide abundant and valuable user-side context information~(e.g., real-time satisfaction feedback, extensive context information, and a clearer description of information needs) to the search engine, which is hard to capture in the previous paradigm.
  In our implementation, the system can decode user satisfaction from brain signals in real-time during the interaction process and re-rank the search results list based on user satisfaction feedback. 
  The demo video is available at \\ http://www.thuir.cn/group/YQLiu/datasets/BMSISystem.mp4~.

\end{abstract}



\maketitle

\section{Introduction}
Adopted in diverse environments and used by billions of users, search engines have changed how humans learn and think. 
Driven by the diversity of information needs and benefiting from the increase in computing resources, search technology is evolving to become more powerful. However, the fundamental interaction paradigm has been relatively stable for decades. 
When searching, a user needs to formulate a query, which often consists of a few keywords, according to his information need, and submit it to the search engine. Upon receiving the query, the search engine will retrieve and return a ranked search results list to users. 
However, there exists several shortcomings of the current search interface: 
1) users tend to issue short queries which bring uncertainty and ambiguity.
Due to the strong dependence on the formulated query, the information loss of bi-directional transmission between users and search engines has caused a significant performance bottleneck~\cite{liu2021challenges}.
2) traditional search systems collect implicit user feedback such as click and dwell time and attempt to connect implicit user feedback with user's subjective feelings. 
But this implicit feedback is usually inaccurate and noisy and may not necessarily align with the subjective feelings of real users.
3) current search interface requires users to interact with search engines using the mouse and keyboard, which is impractical for scenarios where hand-based interactions are infeasible. 

Recently, with the development of \ac{BMI}, it is possible to change the search interface to circumvent the problems mentioned above.
\ac{BMI} provides a direct communication pathway between an enhanced or wired brain and an external device, which is widely applied in researching, mapping, assisting, augmenting, and repairing brain functions.
In the area of non-invasive BMIs, the most popular choice is \ac{EEG}, which has attracted a large amount of theoretical and applied researches in text inputting, Human-Machine interaction, and cognitive activities analysis. 
Taking information transfer rate and visual theoretical study into consideration, \ac{SSVEP} paradigm is applied to implement the module of query inputting and the interaction between humans and the search engine in our system. 
This paradigm assigns each target key~(alphabet or function key) in the virtual keyboard with different flicker frequencies. When the user gazes at certain target key, the SSVEP signal with the same frequency~(as well as its harmonics) is elicited in the visual cortex of the brain. Through analyzing this evoked signal, the system will get the target key which the user intends to enter.  
Besides the ability of system control, brain signals can be decoded as the search context and user feedback to understand and improve the search process.
For example, \citet{moshfeghi2016understanding} find that brain signals are related to the occurrence of information need and \citet{gwizdka2017temporal} decode brain signals to the relevance judgment.

In this paper, we build a search system based on the EEG device.
Users can perform search tasks including formulating/submitting queries, interacting with search results.
Moreover, we estimate the search state and decode user satisfaction and utilize the inferred feedback for evaluating and re-ranking search results to improve search experience. 
As far as we know, our implemented system is the first closed-loop system that users can interact with without relying on the mouse or keyboard.
Based on collected brain signals, the proposed system can improve the search experience proactively, dynamically and personally. 


\section{SYSTEM OVERVIEW}
\label{SYSTEM OVERVIEW}
\begin{figure*}[htbp]
    \centering
  \includegraphics[width=\textwidth]{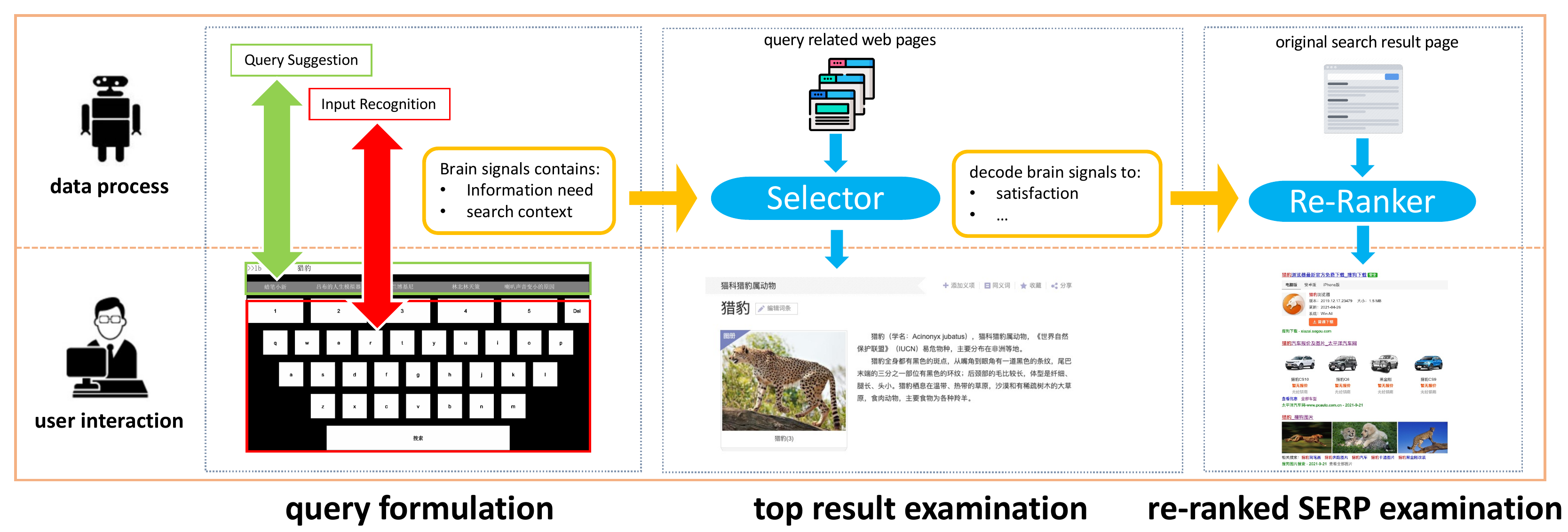}
     \caption{Brain-Machine Search Interface System~(BMSI) consists of the User Interaction Module and the Data Process Module. 
    The User Interaction Module provides the interaction interface including the visual speller page, landing pages, and SERPs. 
    The Data Process Module decodes the brain signals and provide real-time feedback for User Interaction Module.}
    \label{fig:architecture}
\end{figure*}

\ac{BMSI} System consists of two parts: User Interaction Module and Data Process Module, as shown in Figure~\ref{fig:architecture}. 
The User Interaction Module provides interaction interfaces including the visual speller page, landing pages, and SERPs, on which users can perform search tasks such as formulating queries and examining search results.
While the Data Process Module runs in the backend and it can decode the brain signals and provide real-time feedback for the User Interaction Module.

In the visual speller page, users can issue queries by gazing at the target key that flickers with a specific frequency, while the Data Process Module will analyze harmonics evoked in the visual cortex~(nine parietal and occipital channels, including Pz, PO3, PO5, PO4, PO6, POz, O1, Oz, and O2) to locate the target key chosen by the user. 
During the query formulation process, \ac{BMSI} system also records brain signals in other areas of the brain cortex, like the frontal, speech, and reading cortex, which is related to attention, mental status, language understanding, etc. 
The signals in these areas provide rich contextual information about users' information needs and search context, such as the formulation difficulty of summarizing information needs into the query, and users' current status~(working, entertainment or exercise). 
We leave exploring these signals to further boost the search experience as future work.


Once the user finishing the query formulation and clicks the search button, a top-ranked page is presented and attempts to satisfy the users' information need directly, similar to the landing page after the user pressing on the ``I'm Feeling Lucky'' button in Google. 
The selection strategy of the top-ranked page could make use of brain signals in query formulation, which is expected to meet user needs instantly. 
In our demo system, we select the top result in the original \ac{SERP}.
During the examination process of the top-ranked page, the Data Process Module will use brain signals to decode user satisfaction in real-time. 
Then the system would re-rank the search results list according to the detected satisfaction feedback and the user can continue to examine more search results on this re-ranked \ac{SERP}. 
On the re-ranked \ac{SERP}, interaction options including clicking search results, scrolling up or down are given.
These interaction options are provided using several blocks with different flickering frequencies while these blocks are displayed in the right position of the current viewport.
Similar to the interaction paradigm in the visual speller page, search interactions on the re-ranked \ac{SERP} are also based on the evoked SSVEPs and does not need users to use the mouse or keyboard.

\section{APPROACHES}
\subsection{SSVEP based Keyboard}
Neurological research suggests that SSVEP signals are natural responses to visual stimulation at specific frequencies. 
When the retina is excited by a visual stimulus ranging from 3.5 Hz to 75 Hz\cite{beverina2003user}, the visual cortex of the brain generates electrical activity at the same (or multiples of) frequency of the visual stimulus. 
We design a 33-target BMI speller referred to BETA~\cite{liu2020beta} for visual stimulation to evoke SSVEP, and the flicker frequency ranges from 8 to 15.68 Hz. 
To make sure that users can get used to this system easily, we resemble the conventional QWERTY keyboard to construct the graphical interface in which 33 target keys, including 5 numbers, 26 alphabets, and 2 function signs~(Delete and Search), are aligned in five rows. 
Among them, the 5 numbers are used to select candidate words. 

A sampled sinusoidal stimulation method \cite{chen2014high} is adopted to present the visual flicker on the screen. 
In general, the stimulus sequence of each flicker can be generated by 
\begin{equation}
s(f, \phi, i)=\frac{1}{2}\{1+\sin [2 \pi f(\frac{i}{ \text { RefreshRate }})+\phi]\}
\end{equation}
where $i$ denotes the frame index in the stimulus sequence, and $f$ and $\phi$ denote the frequency and phase values of the encoded flicker that uses a joint frequency and phase modulation\cite{chen2015high}. 
The grayscale value of the stimulus sequence ranges from 0 to 1, where 0 indicates dark, and 1 indicates the highest luminance of the screen. For the 33 targets, the tagged frequency and phase values can be respectively obtained by 
\begin{equation}
\begin{array}{c}f_{k}=f_{0}+(k-1) \cdot \Delta f \\ \Phi_{k}=\Phi_{0}+(k-1) \cdot \Delta \Phi\end{array}
\end{equation}
where the frequency interval $\Delta f$ is 0.24 Hz, the phase interval $\Delta \Phi$ is 0.5$\pi$, and k denotes the target index. 
In our work, f0 and $\Phi_{0}$ are set to 8 Hz and 0 $\pi$, respectively. 

\subsection{Input Recognition Algorithm}


By analyzing the evoked SSVEPs, Input Recognition Algorithm could recognize the target key of user intent. 
In our system, \ac{CCA} is applied to measure the canonical correlation coefficient between the real-time SSVEPs and the reference signals, i.e., the theoretical brain signals evoked by the stimulus flickered at a specific frequency. 

Specifically, the SSVEPs can be expressed by the following formula:
\begin{equation}
\mathbf{S} =
\begin{pmatrix}
\mathbf{x_1},\mathbf{x_2},\mathbf{x_3}, \cdots \mathbf{x_9}
\end{pmatrix}^\mathrm{T}
\end{equation}

and the reference signals are:
\begin{equation}\label{signal}
\mathbf{R_f} =
\begin{bmatrix}
sin(2\pi ft)\\
cos(2\pi ft)\\
\vdots\\
sin(2\pi Nft)\\
cos(2\pi Nft)\\
\end{bmatrix},    t=\frac{1}{F_s},\frac{2}{F_s}\cdots \frac{N_s}{F_s} \\
~\\
\end{equation}
where N is the number of harmonics, $f$ is the reference frequency($f=8.00,8.24,8.48 \cdots 15.68$), $F_s$ is the sampling rate and $N_s$ is the number of sampling points. 

For each flicker frequency ranging from 8 to 15.68Hz, we use Eq.(\ref{signal}) to generate the reference signal then calculate the correlation between each reference signal and SSVEP signal. After calculation, the reference signal with the highest correlation is the recognition result, which indicates the target key that the user intends to enter.


\subsection{Query Suggestion}

User search intents are complex.
Hence, a single query is hard to fully express their information needs. 
In that regard, query suggestion techniques can help users to complete their search tasks with less effort in complex search scenarios to some extent.
Especially for Chinese users, the alphabet letters are usually not the final expression of queries but are used to input PinYin and then select the right Chinese words as the query string. 
Under this circumstance, query suggestion is more difficult for Chinese search engines but is strongly necessary to quickly capture users' information needs. 
The function of query suggestion in our system is powered by an API provided by Sogou Inc~\footnote{www.sogou.com}, one of the most popular Chinese search engines. 
The suggestion model integrates the analyses of large-scale heterogeneous data and user behavior on the Internet, such as user clicks, query reformulation, and tracking of hot news, which could effectively bridge the gap between user intent and query candidates.

Query suggestion can provide appropriate query candidates with the incomplete part of a query.
To test the efficiency of the query suggestion model, we simulate the input of 2,956 query samples with page views greater or equal to 500 times in the past half-year.
Our stimulation adopted the two most popular PinYin input strategies: first letter spelling and full letter spelling. 
If query candidates are the fine-grained or same intent of the query sample, we define this as a successful match. 
For example, we regard all these candidates, ``youtube online'', ``youtube download'',  and ``youtube'', as successful matches with the query ``youtube''. 
We show the performance of these two strategies in Table~\ref{tab:input}, we can observe that inputting queries by the first letter require less effort while using full letter spelling can achieve a higher match ratio.

\begin{table}
  \caption{The performance of the query suggestion model with different input strategies.}
  \label{tab:input}
  \begin{tabular}{ccccc}
    \toprule
    Chinese Input Strategies & Successful Match Ratio & \#Keys per Char \\
    \midrule
    first letter spelling & 0.77 & 0.65 \\
    full letter spelling & 0.91 & 1.16 \\
  \bottomrule
\end{tabular}
\end{table}

\subsection{Brain Signals Decoding}

Existing works in multichannel \ac{EEG}-based prediction usually need effective feature extraction. 
Several features are proposed throughout literature and among these features, \ac{DE} is widely used and performs better than other features including band power, rational asymmetry, and differential asymmetry in an multi-channel \ac{EEG}-based emotion recognition task~\cite{duan2013differential}.
Therefore, we extract \ac{DE} features using Short Time Fourier Transform over five frequency bands~(delta: 0.5-4Hz, theta: 4-8Hz, alpha: 8-13Hz, beta: 14-30Hz, gamma: 30-50Hz) in 62 distinct \ac{EEG} channels except for two re-reference channels.
For classification model, we apply \ac{GBDT}, which can automatically choose and combine the \ac{EEG} features and it has shown effective in usefulness estimation with multichannel \ac{EEG} features.
To train the prediction model, we use the Search-Brainwave dataset~\footnote{http://www.thuir.cn/group/YQLiu/datasets/SearchBrainwaveDataset.zip}.
The dataset contains \ac{EEG} data recorded during the 18 participants doing pre-defined search tasks in a period of 60 minutes as well as the usefulness annotation for each search result.
To tune the hyper parameters, including learning rate, estimator number, leaf nodes, and the maximum tree depth, we applied the protocol of leave-one-participant-out.
The protocol means that we apply data of each participant for validation and train the classifier with left participants.
The parameters are tuned according to the averaged validation \ac{AUC} of each participant and then we train our final classifier with all data.
As a result, the system can achieve an averaged \ac{AUC} of 0.69 in validation and the whole steps~(feature extraction and \ac{GBDT}-based classification) described above cost averagely 0.2 seconds in our practice.

Now that we can predict the satisfaction feedback and understand the perceived usefulness of the search results, our search system can automatically adjust to improve the search process.
In our practice, we apply a simple strategy as a first step to utilize brain signals as feedback for a more proactive search system.
Before our experiment, each search result involved has been annotated with some subtopics by topic model. 
On the one hand, when the system notice that the user perceive certain landing page is useful, search results share similar subtopics would be moved to the top.
On the other hand, when we detect that the user is unsatisfied with certain landing page, we will re-rank the search result list and the results share the same subtopics will be moved to the back.
The certain landing page refers to the top-ranked page as described in Section~\ref{SYSTEM OVERVIEW}.
Note that we focus on the effectiveness of brain signals can be decoded as feedback for a better search performance in a real-time system, the methods of how to combine these feedback with other evaluation framework are left as future work.

\section{DEMONSTRATION}


\begin{figure}
  \centering
  \includegraphics[scale=0.41]{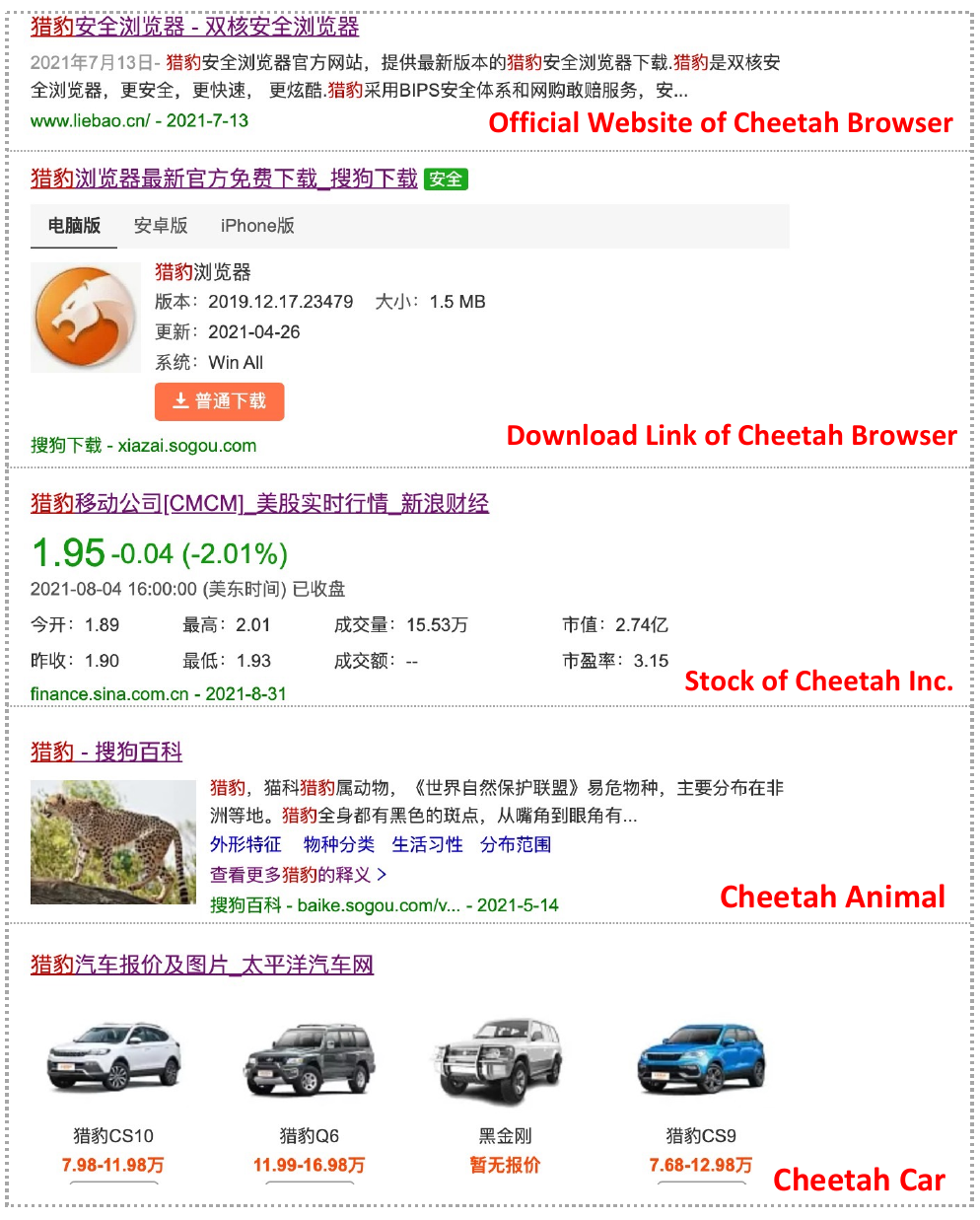}
  \includegraphics[scale=0.41]{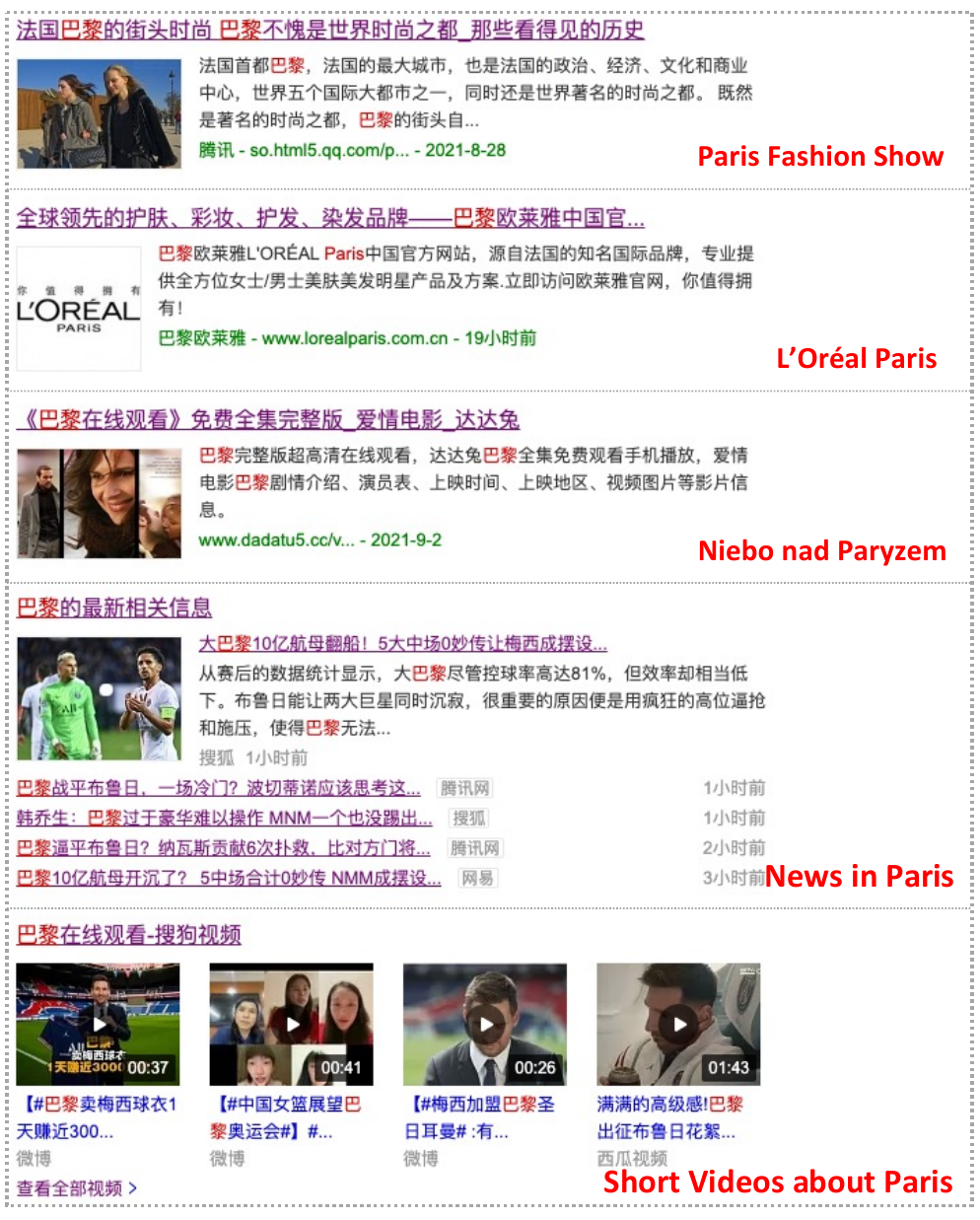}
  \caption{The re-ranked SERPs of two queries are shown above. The user is satisfied with the top-ranked page about Cheetah Browser, so on the left page, the search results related to Cheetah Browser are higher than others. While for the right page, the user is unsatisfied with the city introduction page of Paris, more diverse results are shown at the front of the SERP.}
  \label{fig:serp}
\end{figure}

In this section, we apply two cases to elaborate our \ac{BMSI} System. 
More detailed information is shown in our video.

In our first case, the user wants to learn more about Cheetah~(LieBao in Chinese) Browser, a Web Browser developed by a Chinese company. Then he inputs ``lb'', the first letters of PinYin to ``LieBao''. 
Our system automatically generates a candidate query list with PinYin completion and query suggestion. In this situation, the first candidate meets his information needs, and he can ``press'' the ``search'' button without extra selection. 
Then the system will present the related top-ranked page, which is the official website of Cheetah Browser. During the examination of the top-ranked page, the system decodes the user's satisfaction with brain signals in real-time, and it infers that the user is satisfied with this page. Therefore, when the user continue to examine the \ac{SERP}, as shown on the left of Figure \ref{fig:serp}, the search results related to the subtopic of Cheetah Browser will be ranked higher than others due to our re-ranking strategy.

In our second case, the user plans to download some pictures about the Paris Fashion Show.
The user inputs ``bl'' and selects ``BaLi''(Paris in English) as the query. The presented top-ranked page is a wiki page of the city, which does not meet the user's information need, and the satisfaction decoding module perceives this feedback. 
To find useful information, the user chooses to examine more search results. 
On the re-ranked SERP, the diverse search results related to Paris, such as fashion shows, cosmetics, movies, etc, are provided, while results related to the subtopic that introducing this city are ranked lower.
The corresponding re-ranked SERP is displayed on the right of Figure \ref{fig:serp}.

\section{CONCLUSION}
We develop a web search system via an efficient and effective brain-machine interface. The user inputs query and interacts with the search engine via brain signals only. In the interaction process, the system can collect users' brain signals in real-time, obtain users' feedback by decoding neurological activities, and improve the search experience by query recommendation, search result re-ranking, etc.

\bibliographystyle{ACM-Reference-Format}
\bibliography{sample-base.bib}
\end{document}